\documentclass[aps,prb,reprint,groupedaddress]{revtex4-1}
\usepackage{amsmath,amssymb}

\newcommand{\ket}{\rangle}
\newcommand{\bra}{\langle}

\DeclareMathOperator{\const}{const}
\DeclareMathOperator{\Proj}{Proj}

\newcommand\spinup{\mathord{\uparrow}}
\newcommand\spindown{\mathord{\downarrow}}
\newcommand{\hilb}{\mathcal H}
\newcommand{\C}{\mathbb{C}}

\usepackage{hyperref}
\usepackage{graphicx}
\usepackage{color}

\begin{document}


\title{Geometry of projected connections, Zak phase, and electric polarization}


\author{A.S.~Sergeev}
\affiliation{Faculty of Physics, M.V.~Lomonosov Moscow State University, Moscow, 119991, Russia}


\date{\today}

\begin{abstract}
The concept of the Zak phase lies at the core of the modern theory of electric polarization. It is defined using the components of the Bloch wave functions in a certain basis, which is not captured by the standard expression for Berry potential. We provide a consistent geometric interpretation of the Zak phase in terms of projected connections. In the context of Bloch states, we relate the transformation law of projected Berry potential with classical currents that contribute to the time derivative of the electric polarization. This gives a new argument for the Zak phase formula for the electronic contribution to the polarization. We demonstrate that the Wannier functions play a key role in the description of an adiabatic current in a periodic system.
\end{abstract}

\pacs{71.20.-b, 77.22.Ej, 03.65.Vf}

\maketitle
\paragraph*{Introduction.}

Geometric phases are ubiquitous in physics: in a sense, any gauge field is characterized by the geometric properties of the underlying bundle, such as connection and curvature. Corresponding physical quantities are gauge potential and field strength, the latter being independent of the gauge choice. An important special case of connections are those that originate from projecting a connection in some space to a subspace thereof. Examples of such projected connections include Levi-Civita intrinsic derivative, whose abstract version is important in Riemannian geometry \cite{Frankel2011}, and Berry connection in quantum mechanical systems controlled by time-dependent external parameters \cite{Berry84, Simon83}. In the second example, adiabaticity condition restricts the time-evolved state to a fixed eigenspace of the Hamiltonian, which is a subspace of the Hilbert space of states $\hilb$. 

After a cyclic adiabatic evolution, a quantum state acquires a phase factor, known as the Berry phase.  Remarkably, an explicit time dependence disappears from the description of the process, and the phases are captured only by the geometry of the complex vector bundle over the parameter space.  Therefore, quantum geometric phases need not be related to the time evolution, with the Zak phase as a vivid example \cite{Zak89}. The Zak phase is acquired by a Bloch eigenstate after going around a non-contractible loop in the Brillouin zone (which means taking a contour integral of a potential rather than a physical process). In the framework of the modern theory of polarization\cite{RV-chap}, the Zak phase describes the electronic contribution to the electric polarization of a crystal.

In contrast to the Berry phase, the Zak phase does not have a straightforward geometric interpretation. There are two common basis choices for $\hilb$, which is reflected in Bloch wave function spatial components: $\psi_{k}(x)=e^{ikx}u_{k}(x)$. The Zak phase is defined strictly in terms of $u_{k}$, which leads to controversy since the standard expression for Berry potential ${A=i\bra \psi_k|\partial_k \psi_k\ket}$ leaves the basis choice implicit. Another problem arises because $u_k$ are not periodic as functions of $k$: $u_k\neq u_{k+G}$, where $G$ is a reciprocal lattice vector. Thus the corresponding basis states and the Hamiltonian expression  have a discontinuity at the Brillouin zone (BZ) boundary, which harms the geometric picture of the Berry phase. This forced Zak to define an ``open-path geometric phase'' for a case when the states at the endpoints of an open path are related by a fixed unitary transformation $U$ in $\hilb$ \cite{Zak89a}. In numerical applications, one simply inserts $U$ into the chain of projection operators used to compute the geometric phase\cite{Vanderbilt-BP}. The transformation $U$ is also implicitly present in the Wilson loop formulation\cite{Benalcazar17}  of geometric phase, which is based on the projected position operator\cite{Resta98}  (see Chap.~5 of Ref.~\onlinecite{Resta96} for an exposition).   The two choices of basis in $\hilb$ lead to the definition of two kinds of the Zak phase\cite{Rhim16}, which correspond to different types of the bulk polarization introduced in Ref.~\onlinecite{Watanabe18}. The basis choice is also important in the context of the bulk-boundary correspondence for 1D insulators with inversion symmetry\cite{Rhim16} and in the classification of the chiral symmetric wires\cite{[{See Sec.~VI of Supplemental material in }]Tuegel18}. 

In this work, we give an unambiguous geometric interpretation of the Zak phase and related Berry phases of Bloch states. We also discuss a physical meaning of these phases in terms of the classical bulk currents induced by an adiabatic variation of the crystal Hamiltonian. In the geometric part, we focus on the relation between Berry potential and a basis  choice in $\hilb$. This subtle form of gauge dependence does not appear in the original problem of time evolution considered by Berry and has been pointed out only recently \cite{Fruchart14, Moore17}. We start with a  discussion of the geometric nature of Berry potential, which allows us to define a projected connection and to interpret the Zak phase in these terms. We derive a transformation law for projected Berry potential under basis change in $\hilb$. As for the physical interpretation, we show that the time derivatives of the corresponding Berry phases of Bloch electrons describe the classical adiabatic currents. As an unexpected by-product, this provides a new argument for the relation between the Zak phase and electric polarization. Finally, we consider an illustrative example of an adiabatic charge pump and discuss the role of Wannier functions in determining the current in a periodic system. 

\paragraph*{Projected connections.} 
Let ${L\times\hilb}$ be a complex vector bundle with one-dimensional base space $L$ and a complex $n$-dimensional vector space $\hilb$ as a fiber. Informally, we attach a copy $\hilb_k$ of $\hilb$ to each point $k$ of $L$. Consider a smooth vector field of unit vectors $|\psi_k\ket\in \hilb_k$ over $L$ and corresponding complex lines 
\begin{equation}
V_k^\psi = \{a|\psi_k\ket, a \in \C \}\subset\hilb_k.
\end{equation}
The collection $V^\psi = \{V^\psi_k\}$ can also be thought of as a vector bundle over $L$. Using $|\psi_k\ket$ as a basis for $V_k^\psi$, define a unit vector field 
\begin{equation}
|v\ket = e^{i\phi}|\psi\ket
\end{equation}
with $\phi$ a smooth function of $k$. 

How to define a derivative $\nabla_k|v\ket$ such that the result of differentiation  will belong at each $k$ to the space $V_k^\psi$? One can simply differentiate the scalar component $(\partial_k e^{i\phi})|\psi\ket$, but this is very unnatural: another basis choice will give a different result. Alternatively, we can use a derivative in ambient space:
\begin{equation}\label{DerH}
\partial_k |v\ket = \sum_\alpha (\partial_k v_\alpha) |\alpha\ket,
\end{equation}
where $\alpha$ enumerates basis vectors in $\hilb$; but $\partial_k |v\ket$ need not belong to $V_k^\psi$ in this case. One obtains the desired result by mimicking the definition of the Levi-Civita intrinsic derivative. Define $\nabla_k$ as a composition of $\partial_k$ with projection to the $V_k^\psi$ subspace:
\begin{equation}
\nabla_k |v\ket = \Proj_k^\psi \partial_k |v\ket,
\end{equation}
where $\Proj_k^\psi = |\psi_k\ket\bra\psi_k|$. We shall call $\nabla_k$ a covariant derivative.

To understand the physical significance of $\nabla_k$, consider a parallel transport vector field $|v\ket$, which satisfies $\nabla_k|v\ket=0$. From 
\begin{equation}
0=\nabla_k(e^{i\phi}|\psi\ket) = (i\partial_k\phi+\bra\psi|\partial_k\psi \ket) |v\ket
\end{equation} 
we have
\begin{equation}
|v\ket = \exp \biggl( i\int Adk\biggr)|\psi\ket, \qquad A=i\bra\psi|\partial_k\psi\ket.
\end{equation}
One immediately recognizes in $A$ the gauge potential that describes the Berry phase, once $L$ is identified as a space of control parameters and $|\psi\ket$ as an eigenstate that belongs to the Hilbert space $\hilb$. This is an example of a general relationship between parallel transport and covariant derivative, which are different faces of a mathematical structure called connection on a vector bundle. We conclude that the covariant derivative $\nabla_k$ for $V^\psi$ obtained by projection corresponds to the parallel transport defined by the adiabatic quantum evolution.

Now let us consider the derivative $\partial_k$ for ${L\times\hilb}$ from this perspective. In Eq.~(\ref{DerH}) we have implicitly used that $\partial_k|\alpha\ket=0$, that is, the derivative vanishes on basis vectors in $\hilb$. This is unsettling: we have  just discarded a similar definition for $\nabla_k$ because of the basis dependence. Let $|\alpha^1\ket$ and $|\alpha^2\ket$ be two basis choices, so that 
\begin{equation}
|\alpha^2\ket = \sum_\beta |\beta^1\ket\bra\beta^1|\alpha^2\ket = \sum_\beta|\beta^1\ket U^{12}_{\beta\alpha}.
\end{equation}  
If the transformation matrix $U^{12}$ depends on $k$, the condition $\partial_k|\alpha^j\ket=0$ clearly cannot be simultaneously satisfied for both bases $j=1,2$. We conclude that each basis choice $|\alpha^j\ket$ determines corresponding covariant derivative $\partial^{(j)}$ in ${L\times\hilb}$, such that the basis vectors are parallel transported:
\begin{equation}\label{BasisPT}
\partial^{(j)}|\alpha^j\ket=0.
\end{equation}
Different derivatives $\partial_k^{(j)}$ will give rise to different Berry potentials and curvatures. This gauge dependence does not appear in the case of the adiabatic evolution originally considered  by Berry. For example, in the paradigmatic system of a spin-$\frac{1}{2}$ particle in a rotating magnetic field it is natural to assume that the definitions of the basis vectors $|\spinup\ket, |\spindown\ket$ do not vary in time.  However, this need not be true in other physical context.

This situation is described in Ref.~\onlinecite{Moore17} as a dependence of Berry connection on a trivialization of a Hilbert bundle. By trivialization one means choosing a global section (i.e., a basis in each space $\hilb_k$) that allows one to identify ${L\times\hilb}$ with a trivial bundle ${L\times\C^n}$. As a slight refinement of this approach, note that $|\alpha\ket$ as a definition of parallel transport has more freedom than $|\alpha\ket$ as a trivializing section. Like any other parallel transported vectors, $|\alpha\ket$ need not form a smooth vector field along a closed path in $L$.

Now let us find a transformation law of projected Berry potential upon a basis change in $\hilb$. Note that $\partial_k^{(j)}$ acts on scalars as an ordinary derivative $\partial_k$. Let $\psi^j_\alpha(k)$ denote a component of $|\psi_k\ket$ with respect to a basis $|\alpha^j\ket$. We have
\begin{equation}
\sum_\beta \overline{\psi_\beta^2}\partial_k \psi_\beta^2 = 
\sum_\beta \overline{\psi_\beta^1}\partial_k \psi_\beta^1 + \sum_{\alpha \beta \eta} \overline{\psi_\eta^1}\, U_{\eta \beta}^{12} \, (\partial_k  U_{\beta \alpha}^{21}) \, \psi_\alpha^1 ,
\end{equation}
where the bar denotes complex conjugation. Taking Eq.~(\ref{BasisPT}) into account, one finds
\begin{equation}\label{Transform}
\bra \psi|\partial_k ^{(2)}\psi\ket  = \bra \psi|\partial_k ^{(1)}\psi\ket +
\bra \psi|U^{-1}(\partial_k  U)|\psi\ket,
\end{equation}
where $U=U^{21}$. Reference \onlinecite{Fruchart14} contains a similar expression for Berry curvatures. Note that the local values of geometric quantities do not affect the global topological properties of a bundle. In particular, Chern numbers are independent of the choice of the connection discussed here.

\paragraph*{Geometry of Zak phase.}
Consider a one-dimensional periodic crystal that contains $N$ unit cells with $n_{o}$ atomic orbitals in each cell. Denote $|\alpha_m\ket$ an $\alpha$th orbital within $m$th unit cell. Let $\tau_\alpha$ denote the position of the orbital inside a unit cell, so that the position operator acts on $|\alpha_m\ket$ as
\begin{equation}
\hat x|\alpha_m\ket = (m+\tau_\alpha) |\alpha_m\ket.
\end{equation}
Define the Bloch wave basis as a Fourier transform of real-space orbitals:
\begin{equation}
|\alpha^1_k\ket = \frac{1}{\sqrt{N}} \sum_m e^{imk}|\alpha_m\ket,
\end{equation}
where $k$ is the crystal momentum and the lattice constant is $a=1$. In the basis $|\alpha^1_k\ket$, the Hamiltonian becomes a smooth collection of operators $\hat H_k$ parameterized by $k$. They act on Hilbert spaces $\hilb_k$ that form a rank $n_o$ vector bundle over the BZ. Let $|\psi_k\ket$ be an eigenstate that corresponds to an isolated filled band:
\begin{equation}
|\psi_k\ket=\sum_{\alpha } \psi^1_{\alpha k} |\alpha^1_k\ket.
\end{equation} 

We further define another basis for $\hilb_k$, which is not periodic in BZ, but knows about spatial positions of the atomic orbitals:
\begin{equation}
|\alpha_k^2\ket = e^{ik\tau_\alpha}|\alpha_k^1\ket.
\end{equation} 
The corresponding wave function components $\psi_{\alpha k}^2$ would be traditionally denoted by $u_{\alpha k}$. Finally, we define Wannier states as a Fourier transforms of $|\psi_k\ket$:
\begin{equation}\label{WnDef}
|w^n\ket = \frac{1}{\sqrt{N}}\sum_k e^{-ikn}|\psi_k\ket.
\end{equation}
In one dimension, the state $|w^n\ket$ is exponentially localized around $n$th unit cell \cite{MLWF}. Its spatial components
\begin{equation}\label{WnComp}
w^n_{\alpha m} = \frac{1}{N}\sum_k e^{i(m-n)k}\psi^1_{\alpha k}
\end{equation}
 depend only on the difference $(m-n)$.

We are now in a position to give a geometric interpretation of the Zak phase. Treating $k$ as a continuous variable and using integration by parts, it is not difficult to show that the zeroth Wannier center coordinate is
\begin{equation}\label{WannierCenter}
\bra w^0|\hat x |w^0\ket = \frac{1}{2\pi} \int\limits_{BZ} i \sum_\alpha \overline{\psi^2_{\alpha k}}\,\, \partial_k \psi^2_{\alpha k} dk = \frac{\gamma}{2\pi},
\end{equation}
which is a definition of the Zak phase $\gamma$. We recognize in the integrand the Berry potential 
\begin{equation}
A^{(2)}=i\bra\psi_k|\partial_k^{(2)}\psi_k\ket
\end{equation}
obtained as a projection of the covariant derivative $\partial_k^{(2)}$ for $\text{BZ}\times\hilb$ that vanishes on $|\alpha^2_k\ket$. Note that the usual expression for Berry potential $i\bra u_k|\partial_k u_k\ket$ implicitly assumes that $\partial_k|\alpha^2_k\ket=0$, which immediately leads to a contradiction since $|\alpha^2_k\ket$ are not periodic in the BZ.

\paragraph*{Physical meaning of transformation law.} Let us consider the transformation law (\ref{Transform}) of projected Berry potentials for Bloch eigenstates $|\psi_k\ket$ decomposed with respect to the bases $|\alpha_k^1\ket$, $|\alpha^2_k\ket$. Multiplying (\ref{Transform}) by $\frac{ie}{2\pi}$, where $e$ is an elementary charge, and integrating over the BZ, we have
\begin{equation}\label{BlochTrans}
\frac{e\gamma}{2\pi} =\frac{e}{2\pi} \int\limits_{BZ} A^{(1)}dk +
\frac{e}{2\pi}\int\limits_{BZ} i \bra\psi_k|U^{12}(\partial_k U^{21})|\psi_k\ket dk. 
\end{equation}
The authors of Ref.~\onlinecite{Rhim16} call the value of the integral in the first term the intercellular Zak phase. Using the inverse transform of Eq.~(\ref{WnComp}), which relates $\psi^1_{\alpha k}$ to $w^0_{\alpha m}$ they show that the first term in Eq.~(\ref{BlochTrans}) equals $e\sum_{\alpha m} m|w^0_{\alpha m}|^2$. It is interpreted as the measure of the extra charge accumulated at the ends of the finite chain, other than the classical bound surface charge. However we find another interpretation useful. Consider a 1D crystal with periodic boundary conditions. One can imagine it as a ring that consists of  $N$ unit cells. Let $m$ range in the interval $[-N/2 +1, N/2]$, where for simplicity we assume $N$ to be even. We divide the ring into two regions, one with $m>0$ and the other with $m<0$, excluding the cell $m=N/2$. We claim that the first term of Eq.~(\ref{BlochTrans}) describes a certain local contribution to the difference $\Delta Q$ between the total charges of the two regions. 

\begin{figure}
\includegraphics[width=\columnwidth]{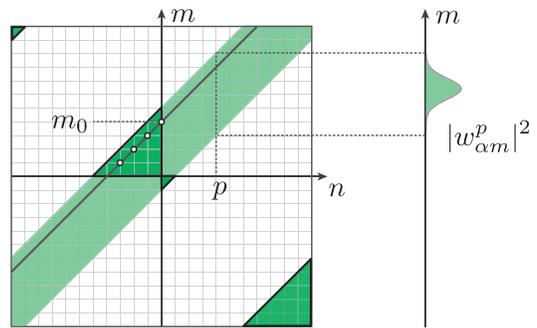}
\caption{Amplitudes of the Wannier states  $|w^n_{\alpha m}|^2$ as functions on the torus parameterized by $(m, n)$. Highlighted triangles contribute to the charge difference between the regions with $m>0$ and $m<0$. \label{Fig:Wannier}  }
\end{figure}

To calculate $\Delta Q$, one can sum the charge density of Bloch functions
\begin{equation}\label{rhoBloch}
\rho_{\alpha m} = \frac{e}{N}\sum_k |\psi_{\alpha k}^1|^2,
\end{equation}
which gives 
\begin{equation}
\Delta Q =\sum_{\alpha , m>0} \rho_{\alpha m} - \sum_{\alpha , m<0} \rho_{\alpha m}=0, 
\end{equation}
since the charge density is cell periodic. It turns out that one can still  extract some physical information from this vanishing difference if the charge density is expressed in terms of Wannier functions $|w^n\ket$. From the definition (\ref{WnDef}) we have $\sum_k \bra \psi_k|\psi_k\ket = \sum_n \bra w^n|w^n\ket$, which yields
\begin{equation}\label{rhoWannier}
\rho_{\alpha m} = e \sum_n|w^n_{\alpha m}|^2.
\end{equation}
Here $n$ enumerates the Wannier functions and has the same periodic range as $m$. We decompose the total charge density as a sum of contributions of individual Wannier functions and plot $|w^n_{\alpha m}|^2$ on the $(m,n)$ torus, as shown in FIG.~\ref{Fig:Wannier}. Each Wannier function $|w^n\ket$ lives on a slice $n=\const$, as shown on the right for $n=p$. The total charge density as a function of $m$ is obtained by adding all the slices together, and further summation in $m$ gives the total charge. Note that different Wannier functions are related by a lattice translation, and the values $|w^n_{\alpha m}|^2$ are constant along the diagonals $m-n=\const$, which follows from Eq.~(\ref{WnComp}). 

We are interested in the difference between the total charges contained in the upper and lower half-planes. Because of the localization, Wannier functions have nonzero components only in some diagonal band. The charges located in the two quadrants with $m\cdot n>0$ cancel each other out, and only the highlighted triangles make a contribution. One pair of the triangles is located near the origin $m=0$ and another one lies near the $m=N/2$ cell, with corresponding contributions to the charge difference $\Delta Q_0$ and $\Delta Q_{N/2} = -\Delta Q_0$. Let us calculate the charge difference contribution $\Delta Q_0$ from the pair located near the origin. We sum over the diagonals specified by the condition $m-n =m_0$. Such a diagonal contains $m_0$ terms, and each one equals $|w^0_{\alpha m_0}|^2$. Upon relabeling $m_0\rightarrow m$, we obtain
\begin{equation}
\Delta Q_0 =  e \sum_{\alpha m} m \big|w^0_{\alpha m}\big|^2,
\end{equation}
which is the expression for the first term on the right in Eq.~(\ref{BlochTrans}).

The second  term in Eq.~(\ref{BlochTrans}) is equal to the dipole moment of the unit cell:
\begin{equation}
\frac{e}{2\pi}\int_{BZ} \sum_\alpha \tau_\alpha |\psi_{\alpha k}^1|^2 dk 
 = \sum_\alpha \tau_\alpha \rho_{\alpha m},
\end{equation}
since for the bases $|\alpha_k^1\ket$, $|\alpha^2_k\ket$
\begin{equation}
(U^{12}\partial_k U^{21})_{\alpha \beta} = -i\tau_\alpha \delta_{\alpha \beta},
\end{equation}
where $\delta_{\alpha \beta }$ is the Kronecker delta.

To restore the dimensional lattice constant, we multiply each term of (\ref{BlochTrans}) by $a$; then we average over the unit cell, so that $a$ cancels out. Finally, we rewrite (\ref{BlochTrans}) as
\begin{equation}\label{QFlow}
e\frac{\gamma}{2\pi} = \Delta Q_0 + \frac{1}{a} \sum_\alpha a \tau_\alpha \rho_{\alpha m}.
\end{equation}

\paragraph*{Classical adiabatic currents.} Next we show how the last equation connects with the modern theory of electric polarization\cite{RV-chap} at the classical level. According to this theory, electric polarization cannot be determined from the knowledge of the charge density $\rho(x)$. The definition of polarization as an average dipole moment of a unit cell  (in one dimension)
\begin{equation}
P_{dip} = \frac{1}{a} \int_{cell} x \rho(x) dx
\end{equation}
 turns out to be inappropriate for several reasons. Instead, the  polarization is defined  as a quantity that satisfies
\begin{equation}
\frac{d P}{dt} = \frac{1}{a}\int_{cell}  j(x, t) dx,
\end{equation}
where $j$ is a microscopic current density induced by an adiabatically slow variation of the state of the crystal. 

To show that these definitions are not equivalent, let us calculate, following Ref.~\onlinecite{Vanderbilt-BP}, the time derivative of the $ P_{dip}$ as a function of the time-dependent charge density $\rho(x, t)$. Using the continuity equation $\partial_t \rho = -\partial_x j$ and integration by parts, one finds:
\begin{equation}\label{ClassFlow}
\frac{1}{a}\int j dx = \frac{1}{a} \bigl(x j\bigr)\big\rvert_0^a +\frac{d}{dt}\biggl(\frac{1}{a} \int x \rho dx\biggr),
\end{equation}
or, equivalently,
\begin{equation}\label{ClassFlowShort}
\dot P = j(a) + \dot P_{dip}.
\end{equation}
Thus the change of the electric polarization results both from the current that flows through the cell boundary  and from the variation of the local dipole moment.

\begin{figure}
\includegraphics[width=\columnwidth]{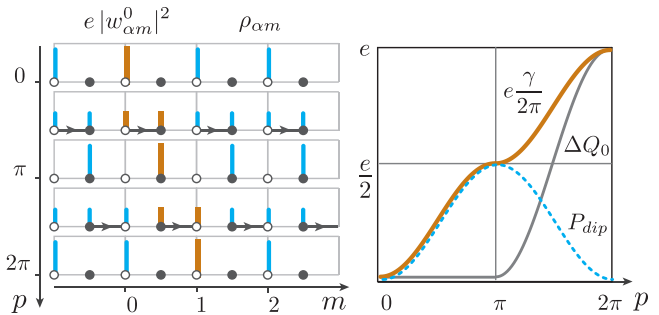}
\caption{ Adiabatic charge pump. Left: evolution of the charge density $\rho_{\alpha m}$ of the diatomic chain as a function of the pumping parameter $p$. Right: terms of Eq.~(\ref{QFlow}) as functions of $p$. The last term is denoted as $P_{dip}$.   \label{Fig:pump}  }
\end{figure}

Now observe that the terms on the right in Eq.~(\ref{ClassFlowShort}) are time derivatives of the right-hand side terms in Eq.~(\ref{QFlow}), understood classically. Indeed, the time derivative of $\Delta Q_0$ is the current flowing through the zeroth unit cell, and the second term is the dipole moment. We conclude that  the Zak phase $\gamma$ describes the electronic contribution to the electric polarization
\begin{equation}\label{ZakPol}
P=e\frac{\gamma}{2\pi}.
\end{equation}
This is the central statement of the modern theory of polarization, which was originally derived by calculating the current in the framework of the linear-response theory \cite{KSV93}. 

\paragraph*{Wannier functions and charge flow.}  Note that the current $\Delta \dot Q_0$ flowing through the cell boundary cannot be calculated from the local charge density (\ref{rhoBloch}), since $\Delta Q_0$ is given by a geometric phase [see Eq.~(\ref{BlochTrans})]. It is crucial to use Wannier functions, whose positions are influenced by the phases of the Bloch states via the Fourier transform. 

To illustrate this point, we consider a simple charge-pumping diatomic chain that switches between an atomic insulator and a chain of disconnected dimers (See Supplemental Material below). The left panel of FIG.~\ref{Fig:pump} shows the evolution of the chain as a function of the pumping parameter $p$. The arrows indicate nonzero hopping amplitudes and the direction of charge pumping due to the variation of the on-site potentials. The bars represent cell-periodic electric charge density $\rho_{\alpha m}$. Orange (darker) bars show the contribution associated with the zeroth Wannier function $e|w^0_{\alpha m}|^2$. 

The values of three terms in Eq.~(\ref{QFlow}) for each $p$ during the pumping cycle are shown in the right panel of FIG.~\ref{Fig:pump}. The heavy line shows the variation of the dipole moment of an elementary charge $e$ placed in the zeroth Wannier center [see Eq.~(\ref{WannierCenter})]. The contribution $\Delta Q_0$ is nonzero only in the second half of the pumping cycle, when the charge transfer between adjacent unit cells occurs. The dashed line shows the change of the local dipole moment of a unit cell, which returns to its initial value after the whole cycle (compare with the evolution of $\rho_{\alpha m}$ for $m=2$ cell in the left panel).

Thus, one can interpret Eq.~(\ref{ClassFlowShort}) quantum mechanically as follows: the position of the Wannier center changes due to the shift of the Wannier function between unit cells and due to the local deformation of the shape of the function. The total charge flow clearly cannot be reconstructed from the charge density $\rho_{\alpha m }$ alone. However, when it is decomposed in terms of Wannier functions [Eq.~(\ref{rhoWannier})], each contribution $e|w^n_{\alpha m}|^2$ contains full information about the current. 

\paragraph*{Conclusions.} In this work, we have discussed geometrical and physical aspects of the Zak phase. We have interpreted the Zak phase geometrically in the framework of projected connections.  We have shown that, for the two common basis choices for Bloch states, the terms in the transformation law for projected Berry potential correspond to the classical adiabatic currents. This result was obtained using the elementary Bloch theory. It gives a new argument for the well-known Zak phase expression for the electric polarization [Eq.~(\ref{ZakPol})]. We have demonstrated that the evolution of each Wannier function contains information about the current through the boundary of the unit cell, a quantity inaccessible from the cell-periodic charge density.

\begin{acknowledgments}
The author thanks  A.J.~Nadtochiy and O.G.~Kharlanov for their valuable comments on the manuscript. Financial support of  RFBR Grants No. 16-29-14037 ofi\_m and No. 16-02-00494 A are acknowledged. 
\end{acknowledgments}

%

\onecolumngrid
\section*{Supplemental Material for: ``Geometry of projected connections, Zak phase, and electric polarization''}
In this Supplemental Material, we define a model of the charge-pumping chain and provide a detailed calculation of all relevant quantities. \\
\twocolumngrid

We consider a diatomic chain with the following Hamiltonian:
\begin{multline}
\hat H = \sum_m (t_{in}|b_m\ket\bra a_m| + t_{ex} |a_{m+1}\ket\bra b_m| +h.c.) +\\+\sum_{\alpha m} U_{\alpha} |\alpha_m\ket\bra\alpha_m|,
\end{multline}
where $|\alpha_m\ket$ denotes the atomic orbital of type $\alpha= a, b$ in the $m$th unit cell, $t_{in}$ and $t_{ex}$ denote hopping amplitudes and $U_\alpha$ stands for the on-site potential. 

In the Bloch basis 
\begin{equation}
|\alpha_k\ket = \frac{1}{\sqrt{N}} \sum_m e^{imk}|\alpha_m\ket
\end{equation}
the Hamiltonian takes the form
\begin{equation}
\hat H = \sum_k \hat H_k = \sum_{k \alpha\beta} |\alpha_k\ket H_k^{\alpha\beta} \bra\beta_k|
\end{equation}
with matrix elements
\begin{equation}
H_k = \begin{pmatrix}
U_a & \overline{t_{in}} +t_{ex} e^{-ik} \\
t_{in} +\overline {t_{ex}} e^{ik} & U_b \\
\end{pmatrix}.
\end{equation}
We assume that hopping amplitudes are real, and on-site potentials satisfy $U_a=-U_b=\Delta$. Then the Hamiltonian matrix can be expressed in terms of Pauli matrices as
\begin{equation}
H_k = \sigma_x (t_{in}+t_{ex}\cos k)  + \sigma_y t_{ex} \sin k  + \sigma_z \Delta .
\end{equation}

Let $p\in[0,2\pi)$ denote the pumping parameter. Define the charge pumping protocol as follows: for the first half $p\in[0,\pi]$ the hopping amplitudes are $t_{in}=\sin p$, $t_{ex}=0$. For the second half $p\in(\pi, 2\pi)$, $t_{in}=0$ and $t_{ex}=-\sin p$. The on-site potential difference is $\Delta = -\cos p$ for all $p$.

For $p\in[0,\pi)$ we have
\begin{equation}\label{Rot1}
H_k =\sigma_x \sin p  -  \sigma_z \cos p,
\end{equation} 
which describes a rotation from the south pole to the north pole of a Bloch sphere in $(x,z)$ plane.

For $p\in (\pi,2\pi)$,
\begin{equation}
H_k = -\sigma_x \sin p \cos k  - \sigma_y \sin p\sin k  -\sigma_z \cos_p ,
\end{equation}   
or, equivalently,
\begin{equation}\label{Rot2}
H_k = \sigma_x \sin p' \cos k +\sigma_y \sin p'\sin k  +\sigma_z\cos p' ,
\end{equation} 
for $p' = p-\pi$, describing a $p'$ rotation about an axis $v_k= (-\sin k, \cos k, 0)$ in the space of Pauli matrices.

Recall that a $\theta$ rotation about an axis $v_\phi = (-\sin \phi, \cos \phi, 0)$ is represented by a unitary transformation 
\begin{equation}
U(\theta, \phi) = \begin{pmatrix}
\cos \frac{\theta}{2} & -\sin\frac{\theta}{2}e^{-i\phi}\\
\sin\frac{\theta}{2}e^{i\phi} & \cos \frac{\theta}{2}\\
\end{pmatrix}
\end{equation}
that relates eigenspinors at two points of the Bloch sphere. 

Consider a low-energy eigenstate $\psi_k^p$ of the $H_k(p)$. Since $H_k(0)=-\sigma_z$, we choose $(1, 0)^T$ as an initial eigenstate. Then the first rotation (\ref{Rot1}) gives
\begin{equation}
\psi_k^{[0,\pi]}=U(-p, 0)\begin{pmatrix}
1 \\0 \\
\end{pmatrix} = 
\begin{pmatrix}
\cos\frac{p}{2} \\ -\sin\frac{p}{2} 
\end{pmatrix}.
\end{equation}
The second rotation (\ref{Rot2}) results in
\begin{equation}
\psi_k^{(\pi,2\pi)} =U(p-\pi, k)\begin{pmatrix}
0\\-1\\
\end{pmatrix} = 
\begin{pmatrix}
-\cos\frac{p}{2} e^{-ik} \\
-\sin \frac{p}{2} \\
\end{pmatrix}.
\end{equation}

The charge density $\rho_{\alpha m}(p) = \frac{e}{N}\sum_k |\psi_{\alpha k}^p|^2$ is found to be  $\rho_{am}(p) = e\cos^2\frac{p}{2}$ and $\rho_{bm}(p) = e\sin^2\frac{p}{2}$ for the whole cycle.

Next we calculate the zeroth Wannier function components $w^0_{\alpha m}$ as Fourier transforms of $\psi_{\alpha k}^p$:
\begin{equation}
w^0_{\alpha m} (p) =\frac{1}{N}\sum_k e^{imk} \psi_{\alpha k}^p.
\end{equation}
For the first half of the cycle,  Fourier transform acts trivially:
\begin{equation}
w^0_m(p) =\begin{pmatrix}
\cos\frac{p}{2}\, \delta_{m, 0}\\ -\sin\frac{p}{2}\, \delta_{m, 0} 
\end{pmatrix}
\end{equation}
and for the second half we have:
\begin{equation}
w^0_m(p)=\begin{pmatrix}
-\cos\frac{p}{2} \, \delta_{m, 1} \\
-\sin \frac{p}{2} \, \delta_{m, 0} \\
\end{pmatrix}.
\end{equation}
 
We place $a$ atom in the origin of a unit cell, $\tau_a=0$, and $b$ atom has the coordinate $\tau_b=\frac{1}{2}$. Then, using the position operator $\hat x |\alpha_m\ket = (m+\tau_\alpha)|\alpha_m\ket$, we calculate the coordinate of the zeroth Wannier center $\bra w^0| \hat x |w^0\ket =\frac{\gamma}{2\pi}$. For the fist half of the cycle,
\begin{equation}
\bra w^0| \hat x |w^0\ket = \frac{1}{2} \sin^2 \frac{p}{2},
\end{equation} 
 and for the second half,
\begin{equation}
\bra w^0| \hat x |w^0\ket = \frac{1}{2} +\frac{1}{2} \cos^2\frac{p}{2}.
\end{equation} 
 
The dipole moment is
\begin{equation}
P_{dip} = e \sum_\alpha \tau_\alpha|\psi_{\alpha k}|^2 = \frac{1}{2}\sin^2\frac{p}{2}.
\end{equation} 

Finally, the charge difference $\Delta Q_0 = e\sum_{\alpha m} m |w^0_{\alpha m}|^2 $ is nonzero only in the second half of the cycle:
\begin{equation}
\Delta Q_0 = e \cos^2\frac{p}{2}.
\end{equation}

\end{document}